\documentclass[12pt]{article}
\usepackage{epsfig}
\usepackage{amsfonts}
\usepackage{amssymb}
\newcommand{\be}{\begin{equation}}
\newcommand{\ee}{\end{equation}}
\newcommand{\beq}{\begin{equation}}
\newcommand{\eeq}{\end{equation}}
\newcommand{\ben}{\begin{displaymath}}
\newcommand{\een}{\end{displaymath}}
\newcommand{\beqa}{\begin{eqnarray}}
\newcommand{\eeqa}{\end{eqnarray}}
\newcommand{\bea}{\begin{eqnarray}}
\newcommand{\eea}{\end{eqnarray}}
\newcommand{\bean}{\begin{eqnarray*}}
\newcommand{\eean}{\end{eqnarray*}}
\newcommand{\ba}{\begin{array}}
\newcommand{\ea}{\end{array}}
\newcommand{\bi}{\begin{itemize}}
\newcommand{\ei}{\end{itemize}}

\newcommand{\ie}{{\it i.e.,\,}}

\newcommand{\bbr}[1]{\mbox{${\mathbb R}^{#1}$}}

\newcommand{\bbz}[1]{\mbox{${\mathbb Z}_{#1}$}}

\newcommand{\reef}[1]{(\ref{#1})}

\begin{document}

\begin{titlepage}
\begin{flushright}
hep-th/0701150
\end{flushright}
\vskip 1.5in
\begin{center}
{\bf\Large{Statistical Description of Rotating Kaluza-Klein Black Holes}}
 \vskip 0.5in {\bf Roberto Emparan$^{a,b}$ and Alessandro Maccarrone$^{b}$}
 \vskip 0.3in {\small
{\textit{$^{a}$Instituci\'o Catalana de Recerca i Estudis
Avan\c cats (ICREA)}}\\
\smallskip
{\it $^{b}$Departament de F{\'\i}sica Fonamental}\\
{\small \textit{Universitat de
Barcelona, Diagonal 647, E-08028 Barcelona,
Spain} }}
\vskip .3 in
{\tt emparan@ub.edu, sandro@ffn.ub.es}
\end{center}
\vskip 0.5in

\baselineskip 16pt
\date{}

\begin{abstract} We extend the recent microscopic analysis of extremal
dyonic Kaluza-Klein (D0-D6) black holes to cover the regime of fast
rotation in addition to slow rotation. Fastly rotating black holes, in
contrast to slow ones, have non-zero angular velocity and possess
ergospheres, so they are more similar to the Kerr black hole. The
D-brane model reproduces their entropy exactly, but the mass gets
renormalized from weak to strong coupling, in agreement with recent
macroscopic analyses of rotating attractors. We discuss how the
existence of the ergosphere and superradiance manifest themselves within
the microscopic model. In addition, we show in full generality how
Myers-Perry black holes are obtained as a limit of Kaluza-Klein black
holes, and discuss the slow and fast rotation regimes and superradiance
in this context.

\end{abstract}
\end{titlepage}
\vfill\eject

\setcounter{equation}{0}

\section{Introduction}

Rotating black holes with maximal angular momentum 
provide an interesting setting for the investigation of
black hole microphysics. Consider the intriguingly simple
form of the entropy of the extremal
Kerr black hole,
\beq\label{kerrs} 
S=2\pi|J|\,.
\eeq 
Since the angular momentum $J$ is naturally quantized, it strongly
suggests that some kind of
sum over states should reproduce it. In addition, the absence of Newton's
constant in \reef{kerrs} gives hope that the counting might be
performed at small gravitational coupling and then reliably extrapolated
to the strong coupling regime where the black hole lies. Identifying the
microscopic system behind \reef{kerrs} remains an open problem in
string theory. This motivates the study of analogous solutions, such as
the extremal Myers-Perry (MP) black hole rotating in two independent
planes in five dimensions \cite{MP}, whose entropy
\beq\label{mps}
S=2\pi \sqrt{|J_1 J_2|}\,
\eeq
is closely similar to \reef{kerrs}, and also of other black holes
sharing some of the features of the Kerr solution.
In this paper we are interested in having an
ergosphere surrounding the horizon.

Recently, a microscopic model for the extremal 5D MP black hole
(orbifolded along a certain direction) has been presented, reproducing
exactly the entropy \reef{mps} \cite{eh}. The model is based on a
connection between the MP solutions and Kaluza-Klein black holes: if we
place an MP black hole at the tip of a Taub-NUT geometry we recover a
Kaluza-Klein black hole. Since Kaluza-Klein black holes are naturally
embedded in Type IIA string theory as solutions with D0 and D6 charges,
ref.~\cite{eh} used the analysis of D0-D6 bound states in \cite{wati} to derive
a microscopic model for \reef{mps}.

Kaluza-Klein black holes are also of interest by themselves. In the
generic dyonic case they are never supersymmetric, nor are in general
close to any supersymmetric state. The entropy of the extremal
solutions---with degenerate horizons of zero temperature---depends, like
\reef{kerrs}, only on integer-quantized charges, and not on the coupling
or other moduli. There are two branches of extremal black holes,
depending on whether their angular momentum is below or above a certain
bound \cite{rash,larsen1}. Ref.~\cite{eh} developed the statistical
description of KK black holes in the slow-rotation regime. In this paper
we extend the analysis to show that the entropy of fastly-rotating KK
black holes can also be accurately reproduced. This is of interest for
several reasons. Unlike the slowly-rotating KK black holes, whose
horizons are static, the fastly-rotating black holes have non-zero
horizon angular velocity, possess ergospheres and exhibit superradiance,
so they are qualitatively much closer to the Kerr black hole. In fact,
as $J$ grows large with fixed charges, the KK black holes asymptotically
approach the Kerr solution. So one may hope for hints for a statistical
model of \reef{kerrs}. 

One feature that we find, and which we argue can be expected for the
extremal Kerr black hole too, is that the microscopic calculations
exactly match the entropy but {\it not the mass} of the fastly rotating
black hole. As we will see, this fits well with the macroscopic analyses of
\cite{astefanesei,dst,astefanesei2} in the context of the attractor
mechanism. Our study also gives a clear indication of how extremal
rotating black holes with superradiant ergospheres are distinguished
microscopically from those that cannot superradiate. Another, perhaps
surprising, consequence of our analysis is that both slowly and
fastly-rotating KK black holes provide microscopic accounts of the
entropy formula \reef{mps}, even if they correspond to rather different
microscopic states. As we discuss below, this does not pose any problem,
since the microscopic theory always retains a memory of how the 5D black
hole is embedded within Taub-NUT.

\setcounter{equation}{0}

\section{Extremal Kaluza-Klein black holes}
\label{sec:KKbhs}

Refs.~\cite{rash} and \cite{larsen1} independently constructed the
solutions for general Kaluza-Klein dyonic rotating black holes, which we give in
appendix \ref{app:MPKK}. For more details we refer to these papers and to
ref.~\cite{larsen2}, which contains insightful remarks on their
properties. Here we briefly summarize the most relevant features.

The solutions are characterized by four physical parameters: mass $M$,
angular momentum $J$, and electric and magnetic charges $Q$ and $P$. For
solutions with a regular horizon, the mass
always satisfies
\beq\label{mgbound}
2G_4 M\geq \left(Q^{2/3}+P^{2/3}\right)^{3/2}\,.
\eeq
The extremal limit, defined as the limit
of degenerate, zero-temperature
horizon, can be achieved in two ways, giving two distinct
branches of solutions:

\begin{itemize}
\item{\bf Slow rotation:} $G_4 |J|< |PQ|$. The mass 
\beq\label{mbound}
2G_4 M= \left(Q^{2/3}+P^{2/3}\right)^{3/2}\,
\eeq
saturates the bound
\reef{mgbound} {\it independently of $J$}. The angular velocity of the
horizon vanishes, and there is no ergosphere. The entropy is
\beq\label{slows0}
S=2\pi\sqrt{\frac{P^2Q^2}{G_4^2}-J^2}\,.
\eeq

\item{\bf Fast rotation:} $G_4 |J|> |PQ|$. The entropy 
\beq\label{fasts0}
S=2\pi\sqrt{J^2-\frac{P^2Q^2}{G_4^2}}\,
\eeq
is the natural continuation of \reef{slows0}, but the mass is strictly
above the value \reef{mbound} and, for
fixed $Q$ and $P$, it grows
monotonically with $|J|$.
The angular velocity of the horizon is non-zero, and there is an
ergosphere. 

\end{itemize}

The extremal horizon disappears and becomes a naked singularity at the
dividing value $G_4 |J|= |PQ|$.

Kaluza-Klein black holes are naturally embedded in Type
IIA string theory by taking a product with $T^6$. The KK gauge potential
is then identified with the RR 1-form potential (so the
KK circle is identified with the M theory direction). $Q$ and $P$
correspond to D0 and D6 charges, quantized as
\beq\label{d0d6}
Q=\frac{g}{4 V_6}N_0 \,,\qquad P=\frac{g}{4}N_6\,,
\eeq
where $N_0$ and $N_6$ are the number of D0 and D6 branes, $g$ is the string
coupling, and the volume of $T^6$
is $(2\pi)^6 V_6$. We work in string units,
so $G_4=g^2/8V_6$. 

The mass bound
\reef{mbound} becomes
\beq\label{slowm1}
M=\frac{1}{g}\left(N_0^{2/3}+(N_6 V_6)^{2/3}\right)^{3/2}\,,
\eeq
and the entropies \reef{slows0} and \reef{fasts0} become
\beq\label{slows1}
S=2\pi\sqrt{\frac{N_0^2N_6^2}{4}-J^2}\,,
\eeq
and
\beq\label{fasts1}
S=2\pi\sqrt{J^2-\frac{N_0^2N_6^2}{4}}\,,
\eeq
respectively. In analogy with \reef{kerrs}, these entropies are independent of $g$,
$V_6$ and any
other $T^6$ moduli.

\setcounter{equation}{0}

\section{Microscopic model of rotating D0-D6 black holes}

The microscopic description of two-charge D0-D6
systems requires that we recall first some aspects of 
four-charge configurations in Type II string theory compactified on
$T^6$ (or M theory on $T^7$), in particular when rotation in the
non-compact directions is present.

\subsection{Rotating zero-temperature configurations in the $(4,0)$-SCFT}
\label{sec:micro1}

Consider brane intersections with four charges $N_1$, $N_2$, $N_3$,
$N_4$, in a regime where the dynamics of low energy modes localized at
the intersection is described in terms of a chiral
$(4,0)$-supersymmetric CFT \cite{Maldacena:1996gb,hlm,msw}. 
We shall be somewhat unspecific about what the $N_i$ stand for. The
statistical entropy counting is most easily performed when $N_{1,2,3}$
denote wrapping numbers of M5 branes, and $N_4$ denotes momentum units
along the (smoothed) intersection \cite{msw}. However, the U-dual
frame where the $N_i$ correspond to four stacks of D3 branes
intersecting over a point \cite{Balasubramanian:1996rx} will be more
useful later. Like in \cite{eh}, the modular invariance of the entropy
and angular momentum makes it natural to assume that the degrees of
freedom responsible for them are localized at the point-like intersection. 

To recover the SCFT we take the number of antibranes of the 1,2,3 kind
to be suppressed, but we allow for both branes (or momentum) and
antibranes (or oppositely moving momentum) of type $4$. To leading order
with $N_{1,2,3}\gg 1$, the central charge for both left- and
right-moving sectors is $c=6N_1 N_2 N_3$, and $L_0-\bar L_0=N_4$.
Supersymmetric configurations
have the left-moving sector in its supersymmetric
ground state, $\bar L_0=N_L=0$ \cite{Maldacena:1996gb}.
We are, however, interested in exciting the left
sector, thus breaking all supersymmetries. The reason is that
spacetime rotation requires exciting
the fermions in this sector. Their $SU(2)$ R-charge acts on
spacetime as $SO(3)$ rotation, so a macroscopic angular momentum $J$
results from the coherent polarization of these fermions. This projection also
reduces the available phase space, so
the effective oscillator number
entering the entropy formula is $\tilde
N_L=N_L-6J^2/c=N_L-J^2/N_1N_2N_3$. Then \cite{hlm}
\beq
S=2\pi\sqrt{\frac{c}{6}}\left(\sqrt{\tilde N_L}+\sqrt{N_R}\right)
=2\pi\left(\sqrt{N_1 N_2 N_3 N_L-J^2} +\sqrt{
N_1 N_2 N_3 N_R}\right)\,.
\eeq

Under the assumption that the
constituents interact only very weakly, the total mass of the system is 
\beq\label{masscft}
M= M_1 N_1+M_2 N_2 +M_3 N_3 +M_4(N_R+N_L)\,.
\eeq
Here $M_i$ are the masses of a unit of each single constituent.

Zero-temperature states must have oscillator distributions such that
either the left or right `temperatures', $T_L$ or $T_R$, vanish. For a
state with $J\neq 0$, this results into two distinct
possibilities:

\begin{itemize}
\item
$T_R=0$

Set $N_R=0$ and
$N_L>\frac{J^2}{N_1 N_2 N_3}\geq 0$, so $N_4=-N_L<0$.
The
left-moving sector gives
rise to both the angular momentum, 
\beq
J^2<N_1 N_2 N_3 |N_4|\,,
\eeq
and the entropy,
\beq\label{slows}
S=2\pi\sqrt{N_1 N_2 N_3 |N_4| -J^2}\,.
\eeq
Hence the mass 
\beq\label{slowm}
M= M_1 N_1+M_2 N_2  +M_3 N_3+M_4 |N_4|\,,
\eeq
is fixed by the
charges $N_i$ independently of $J$. 

\item
$T_L=0$ 

Set $N_R>0$, $N_L=\frac{J^2}{N_1 N_2 N_3}$, so
$N_4=N_R-\frac{J^2}{N_1 N_2
N_3}$. 
The fermions in the left sector fill up to the Fermi level, so 
$T_L$ is effectively zero. 
{\it Both} sectors are
excited, and in principle $N_4$ can be either positive, negative,
or zero. However, if we require that the right sector be only slightly
excited, $N_R\ll N_L$, 
then $N_4<0$. The left movers
provide the angular momentum
\beq
J^2=N_1 N_2 N_3 N_L >N_1 N_2 N_3 |N_4|\,,
\eeq
and the right movers the entropy, 
\beq\label{fasts}
S=2\pi\sqrt{N_1 N_2 N_3 N_R}=2\pi\sqrt{J^2-N_1 N_2 N_3 |N_4|}\,.
\eeq
{}From \reef{masscft} we find the mass
\beqa\label{fastm}
M&=&M_1 N_1+M_2 N_2 +M_3 N_3 
+M_4\left(N_4+2\frac{J^2}{N_1 N_2
N_3}\right)\nonumber\\
&=&M_1 N_1+M_2 N_2 +M_3 N_3 +M_4|N_4|
+2M_4\left(\frac{J^2}{N_1 N_2
N_3}-|N_4|\right)\,
\eeqa
is strictly above \reef{slowm}. 

\end{itemize}

This CFT describes the four-dimensional black holes of \cite{cvyo}
(\cite{hlm}). The restriction to small $N_R$, so that $\frac{J^2}{N_1
N_2 N_3}-|N_4|$ is small, is required for the validity of the CFT
description. Indeed, it is only in this regime that the supergravity
solutions have a locally-AdS$_3$ near-horizon geometry. However, the
entropy of the extremal $T_L=0$ black holes appears to be correctly
reproduced for arbitrary values of the parameters. We will comment more
on this in the final section.

\subsection{Microscopics of D0-D6}

According to \cite{wati}, a system of $N_0$ D0 branes bound to $N_6$ D6
branes wrapped on $T^6$ is T-dual to a non-supersymmetric
intersection of four stacks of D3
branes. One of the stacks has reversed orientation relative to the
supersymmetric case. This is similar to the configurations of the
previous section with $N_4<0$ (the susy-breaking case), but there is one
important difference: the
D3 branes wrap now non-minimal rational directions
$k/l$ in each $T^2$ within $T^6=T^2\times T^2\times T^2$. 
The number $N$ of D3 branes is
the same in each stack,
and 
\beq
N_0=4k^3 N\,,\qquad N_6=4 l^3 N\,
\eeq
(so $N_{0,6}$ are necessarily multiples of four). 

The main assumption of the model, supported by modular invariance, is
that the entropy of the low energy
excitations at the D3 brane intersection is a local property of the
intersection and is independent of whether the branes wrap the torus
along minimal or along non-minimal rational cycles. Then we can import
the entropy calculations from the previous section by setting
\beq
N_1=N_2=N_3=|N_4|=N\,. 
\eeq
Crucially, we must also take into account that the {\it number} of
intersection points in the torus {\it does} depend on how the branes are
wrapped: there are $2kl$ intersections in each $T^2$, and so a total of
$(2kl)^3$ in $T^6$. Since the Hilbert space at each intersection is
independent of the other intersections, the total entropy is $(2kl)^3$
times the entropy from a single intersection point. The angular momentum
is also multiplied by this same factor. Since the total entropy is maximized
by distributing $J$ evenly over all intersections, each one carries
$J_0=J/(2kl)^3$. 

In order to obtain the masses for the
D3-brane configuration we note that if the volume of minimal 3-cycles in $T^6$
is $V_3$, then each of the 3-branes has volume
$(k^2+l^2)^{3/2} V_3$, so their individual masses are 
\beq\label{massd3}
M_{D3}= (k^2+l^2)^{3/2}\frac{V_3}{g}\,,
\eeq
for branes in any of the four stacks.

In this set up, we find that the two different extremal rotating D0-D6
systems of section \ref{sec:KKbhs} map to each of the two
zero-temperature rotating intersecting D3-brane systems of section
\ref{sec:micro1}:

\begin{itemize}
\item {\bf Slow rotation.} This was the regime studied in \cite{eh}. The
CFT at the intersection has the right sector in its ground state, with
entropy per intersection given by \reef{slows}.
So, for the D0-D6 system,
\beq
S_{branes}=(2kl)^3\times 2\pi \sqrt{N^4-J_0^2}=2\pi\sqrt{\frac{N_0^2N_6^2}{4}-J^2}\,,
\eeq
in exact agreement with \reef{slows1}. The mass also 
matches exactly.
Putting $N$ 3-branes with mass \reef{massd3} in each of the four stacks,
the total mass is
\beq
M_{branes}=4N M_{D3}=\frac{V_3}{g}\left(N_0^{2/3}+N_6^{2/3}\right)^{3/2}\,.
\eeq
After T-duality in the three appropriate torus directions, the agreement
with the mass of the slow-rotation D0-D6 black hole \reef{slowm1}, is exact.

\item {\bf Fast rotation.} We naturally
assign to each intersection a state in the fastly-rotating regime of
the CFT, \ie $T_L=0$. Then, using the entropy formula
\reef{fasts},
\beq
S_{branes}=(2kl)^3\times 2\pi \sqrt{J_0^2-N^4}=2\pi\sqrt{J^2-\frac{N_0^2N_6^2}{4}}\,
\eeq
we recover the correct value for the D0-D6 black hole \reef{fasts1}.

The agreement,
however, does not extend to the mass in this case. Consider values of $|J|$
slightly above $N_0N_6/2$, so there is only a small mass $\delta M>0$ above \reef{slowm1},
\beq
M=\frac{V_3}{g}\left(N_0^{2/3}+N_6^{2/3}\right)^{3/2} +\delta M\,.
\eeq
We compute first $\delta M$ within the
microscopic brane model. Recall that the mass is simply proportional to
the volume of branes of each kind, so we use \reef{massd3} in \reef{fastm}. $\delta
M$ comes from the last term
in \reef{fastm}, and we find\footnote{Note that we saturate
$N_L=J_0^2/N^3$ at each intersection, which is smaller than $J^2/N^3$.}
\beqa
\delta M_{branes}&=&\frac{V_3}{g} (k^2+l^2)^{3/2}\times 2\left(\frac{J_0^2}{N^3} -
N\right)\nonumber\\
&=& \frac{V_3}{g}\frac{\left(N_0^{2/3}+N_6^{2/3}\right)^{3/2}}{2}
\left(
\frac{4J^2}{(N_0N_6)^2}-1
\right)\,.
\eeqa
On the other hand, the ADM mass of the black hole gives, after
T-duality, and to leading
order in $(J^2-(N_0N_6)^2/4)$, 
\beq
\delta M_{bh}=\frac{V_3}{g}\frac{(N_0N_6)^{2/3}}{2\left( N_0^{2/3}+N_6^{2/3}\right)^{1/2}}
\left(\frac{4J^2}{(N_0N_6)^2}-1\right)\,.
\eeq
So
\beq
\frac{\delta M_{branes}}{\delta M_{bh}}
=\left[\left(\frac{N_0}{N_6}\right)^{1/3}
+\left(\frac{N_6}{N_0}\right)^{1/3}\right]^{2}\,.
\eeq
The discrepancy in the masses is naturally attributed to a mass
renormalization as the gravitational coupling is increased. Observe that
$\delta M_{branes}>\delta M_{bh}$, which is as expected since
gravitational binding should reduce the energy. In the final section we
discuss further why this
renormalization occurs for fast but not for slow rotation.

\end{itemize}

Following the last comments in the previous subsection, in principle it
would seem possible to extend the agreement of the entropies to
arbitrarily large values of $J^2/N_0^2N_6^2$, but in these cases the use of
the CFT seems largely unjustified. The mass renormalizations get of
course much larger.

\section{Ergospheres and Superradiance}

\subsection{Qualitative microscopics}

Extremal rotating black holes with ergospheres provide a clean setting
for analyzing superradiance. Since these black holes are at zero
temperature, Hawking radiation, which typically mingles with
superradiance, is absent. The distinction between the two effects,
however, is not sharp: the extremal Kerr black hole does spontaneously
emit superradiant modes through quantum effects \cite{page}. This effect
drives the black hole from the extremal to a non-extremal state, and
superradiant emission smoothly mixes with Hawking radiation. 

The statistical description above gives some clear hints of what is the
microscopic distinction between extremal rotating black holes with or
without ergospheres, and how superradiance arises from the microscopic
theory\footnote{The following applies not only to Kaluza-Klein black
holes but also to the four-charge 4D and three-charge 5D black holes,
for which there also exist extremal rotating states with and without
ergospheres. Note also that this is quite independent of the presence or
absence of unbroken supersymmetry.}. Recall first how Hawking radiation
appears microscopically. In the 2D CFT, non-extremal, finite-temperature
states occur when the effective temperature of the excitations in both
left and right sectors is non-zero, with the total system at temperature
$T_H^{-1}=(T_L^{-1}+T_R^{-1})/2$. So if we couple the CFT to closed
strings that propagate out to the asymptotically flat bulk, then left-
and right-moving open string excitations can combine into a closed
string, resulting into Hawking radiation at temperature $T_H$. If
rotation is present, superradiance effects will mix in. However, when
one of the sectors is at zero temperature, \ie at extremality, thermal
Hawking radiation cannot occur. Above we have described two distinct
rotating zero-temperature systems. In the first possibility the
right-moving sector remains unexcited. So, in the absence of open string
excitations of one chirality, there cannot be any closed string
emission---neither Hawking emission nor superradiance. This is as it
should be, since these states describe extremal black holes without
ergospheres.

In contrast, extremal black holes with a superradiating ergosphere
correspond to states with both left- and right-moving excitations. The
emission of (non-thermal) closed strings, from the combination of left-
and right-moving open string excitations, seems possible now. Moreover,
since the left-moving excitations have spin, the emitted closed string
will necessarily carry angular momentum away from the black hole. So it
is natural to expect that this process describes superradiant emission.
The details of this correspondence are being investigated and we hope to
report on them elsewhere.

Let us also mention that a different system where Hawking radiation is
absent but ergoregions exist is provided by the horizonless smooth
rotating solitons of \cite{jejjala}. The fact that these are
dual to CFT states\footnote{In this case, the $(4,4)$-SCFT of the
D1-D5 system.} where both the left and right sectors are excited above
their Ramond ground states is in agreement with the picture we suggest.
These solutions might also provide a convenient setting for the
microscopic study of ergoregions.

\subsection{4D vs 5D perspectives}

The charges $Q$ and $P$, or alternatively the corresponding integers
$N_0$ and $N_6$, have a neat geometrical interpretation from the
five-dimensional point of view: $N_0$ is the number of units of
quantized momentum in the compact fifth direction, and $N_6$ is the
degree of the fibration of this internal $S^1$ on the orbital $S^2$'s.
So when $N_6> 0$ the horizon topology in 5D is $S^3/\bbz{N_6}$. If the
horizon size is much smaller than the compact radius, the black hole can
be regarded as an MP black hole at the tip of a Taub-NUT geometry
\cite{itz}. We elaborate in detail on this in appendix \ref{app:MPKK},
and mention here only some salient features. 

The MP black hole generically has angular momenta
$J_1$, $J_2$ in two independent rotation planes. The KK electric charge
$Q$ is proportional to the self-dual component of the angular momentum,
${\cal J}=J_1+J_2$, aligned with the KK fiber, and $J$ to the
anti-self-dual component $\bar{\cal J}=J_1-J_2$, off the KK direction.
In the extremal limit, the MP black hole entropy reduces to \reef{mps},
which can be written as
\beq
S=\pi\sqrt{|{\cal J}^2-\bar{\cal J}^2|}\,.
\eeq


Both the slowly and the fastly rotating extremal KK black holes lead, in
the decompactification limit, to extremal MP black holes, the former
with ${\cal J}^2>\bar{\cal J}^2$, the latter with ${\cal J}^2<\bar{\cal
J}^2$. From a purely 5D (decompactified) viewpoint, this distinction is
obviously arbitrary. However, we have found that the brane
configurations describing each of these two regimes are rather
different. The point is that the symmetry between $\cal J$ and $\bar
{\cal J}$ is broken once we put the MP black hole at a certain
orientation within Taub-NUT. There is a choice to be made of which of
the 5D angular momenta is going to correspond to the four-dimensional
$J$ and which to $Q$. So the two microscopic configurations actually
describe two different ways to embed the extremal MP black holes within
Taub-NUT, and in this sense they describe different black holes.

One might then ask how it can be that the MP black hole in Taub-NUT is
capable of superradiating when ${\cal J}^2<\bar{\cal J}^2$, but not when
${\cal J}^2>\bar{\cal J}^2$. It turns out that, as we show in detail in
appendix \ref{app:ergo}, superradiance {\it is} possible in both situations but is
interpreted differently in each case. Consider an incident wave in the
KK black hole background with dependence 
\beq
\Psi \sim e^{iky+i n\phi
- i \omega t}
\eeq 
on the Killing directions, $y$ being the coordinate along
the KK circle. The wavenumber $k$ is KK electric charge from
the 4D viewpoint. We find that the necessary condition for superradiant
amplification is
\beq
k<\omega< n\Omega_H + 2G_4k\,\Phi_{E}
\eeq
where $\Omega_H$ is the 4D horizon angular velocity and $\Phi_E$ is the
KK electric potential. From the 5D viewpoint,
$2G_4\Phi_{E}$ is the velocity at which the 5D horizon is rotating in the
$y$ direction relative to static asymptotic observers. 

The conventional rotational superradiance of fastly-rotating KK black
holes corresponds to amplification of neutral ($k=0$) waves with
$\omega< n\Omega_H$. We show in appendix \ref{app:ergo} that this is indeed
possible for scalar fields. This is the process whose microscopic dual
is suggested in the previous section.

On the other hand, slowly spinning extremal black holes have
$\Omega_H=0$ so they show no rotational superradiance. But they can
produce superradiant amplification of waves with KK electric charge $k$.
In appendix \ref{app:ergo} we show that this indeed happens and is always allowed
since these black holes have $2G_4\Phi_{E}>1$. This process, however, is
not so naturally described in the dual CFT system, since it requires
either the emission of 4D charge and hence changing the central charge of the
CFT, or altering the direction in which the branes wrap $T^6$, which is
not seen by the CFT.

\setcounter{equation}{0}

\section{Discussion}

We have shown that it is possible to successfully extend the microscopic
model of KK black holes in \cite{eh} to cover the regime of fast
rotation, with horizons that rotate with non-vanishing angular velocity
and therefore are more similar to the Kerr black hole. There exist other
similar instances where the entropy is also correctly reproduced:
extremal four-charge type II black holes also have slow and fast
rotation regimes which are correctly captured by the CFT of section
\ref{sec:micro1}, and there are analogous three-charge five-dimensional
black holes with these properties which can be described in the
$(4,4)$-SCFT of the D1-D5 system \cite{cvyo,hlm}. However, in these
cases not only the entropy but also the mass is accurately reproduced by
the microscopic model, both at slow and at fast rotation (at least for
rotation slightly above the divide). This agreement is understood,
within the context of AdS/CFT duality, as being due to the existence of
a locally AdS$_3$ (BTZ) geometry near the horizon \cite{cvelar,revper}.
In contrast, extremal KK black holes do not have in general AdS$_3$
symmetry near the horizon (only in the singular case $|PQ|=G_4|J|$, \ie
when either $J_1$ or $J_2$ vanish, do they have in 5D the near-horizon
symmetry $SO(2,2)$ of AdS$_3$ \cite{bh}\footnote{This seems to be
related to the fact that the solution can be reached as a limit of a
black ring \cite{rings}. It is also analogous to the phenomenon
discussed for the BMPV black hole in \cite{natxo,guica}.}), so perhaps
we should be surprised by the fact that the entropy does come out
correctly.

Actually, our results are in perfect agreement with the recent
macroscopic studies in \cite{astefanesei,dst,astefanesei2}, which argue
that the $SL(2,\bbr{})\times U(1)$ symmetry near the horizon of
four-dimensional extremal rotating black holes, charged as well as
neutral, ensures that the macroscopic value for the entropy can be
extrapolated to weak coupling. Extremal KK black holes do possess this
near-horizon symmetry (and their MP limits too \cite{bh}). Hence, if a
microscopic model is identified, its entropy should exactly match the
macroscopic value. In this paper we have provided this microscopic model
and confirmed the agreement of entropies. 

Ref.~\cite{astefanesei} finds that the scalar field, and indeed the
whole solution, for slowly rotating KK black holes is attracted to a
completely fixed form near the horizon, so not only the entropy but also
the mass is fixed---in agreement with the microscopic calculation in
\cite{eh}. However, for fastly rotating extremal KK black holes there
exist flat directions in the effective potential for the scalar near the
horizon, with the effect that only the entropy is attracted to a fixed
value. Other quantities, like the mass, are not guaranteed to be fixed.
We have found that in fact {\it they are not}. Thus we conclude that the
attractor mechanism correctly predicts which quantities will match at
weak and strong coupling, and which ones will, generically, be
renormalized. 

Extremal fastly-rotating four-charge black holes with $J^2\gg
N_1N_2N_3|N_4|$, and KK black holes with $J^2\gg N_0^2N_6^2$, do have
only $SL(2,\bbr{})\times U(1)$ near-horizon symmetry. In principle these
black holes can approach arbitrarily closely to the extremal Kerr
solution. Their entropies, but not their masses, agree with naive CFT
formulas, although one is far from the regime where any application of
the CFT is justified. So, even if this may not be the correct
description, it seems likely that a microscopic model for the extremal
Kerr solution, which also has near-horizon symmetry $SL(2,\bbr{})\times
U(1)$ \cite{bh}, should be able to pin down exactly the entropy
\reef{kerrs}, but not the mass of the black hole. Obtaining the exact
entropy of non-extremal vacuum black holes, like Schwarzschild, will
require taking into account mass renormalization effects.

 \medskip
\section*{Acknowledgements}
\noindent
We thank Gary Horowitz for comments on an earlier version of the
manuscript. This work was supported in part by DURSI 2005
SGR 00082, CICYT FPA 2004-04582-C02-02 and EC FP6
program MRTN-CT-2004-005104. AM was partially supported by a FPU
grant from MEC (Spain).


\medskip

\section*{Appendices}

\appendix

\setcounter{equation}{0}
\section{Myers-Perry from Kaluza-Klein: General case}
\label{app:MPKK}

{}From a five-dimensional standpoint, KK black holes with non-zero
magnetic charge can be regarded as black holes sitting at the tip of a
Taub-NUT space, at least as long as the black hole size is much smaller than
the compact radius. This was noted, in a particular case, in
ref.~\cite{itz}, which showed that in a
limit of large fifth-dimensional radius the {\it static} ($J=0$) dyonic
Kaluza-Klein black holes reduce to five-dimensional Myers-Perry black
holes with {\it self-dual} angular momentum. This corresponds to the case
where the rotation of the MP black hole is aligned exactly along the
Kaluza-Klein direction. In this appendix we consider the most general
case: the KK black hole has non-vanishing four-dimensional angular
momentum $J$, which corresponds to the anti-self-dual component of the
angular momentum of the MP black hole. We keep parameters in
the KK solution arbitrary, in particular we do not confine ourselves to
extremal limits. Hence we are able to recover the general MP solution.

\subsection{Limiting procedure}

We write the solution in essentially the form given in \cite{larsen1},
and use the results therein for the physical parameters\footnote{Here,
and in \reef{eq:general_MP} below, our sign choices for the rotation
parameters are such that positive $\alpha$, $a$, and $b$ correspond to positive
rotation.}.
In five-dimensional form,
\begin{equation}\label{eq:general_KK}
ds^{2}=
\frac{H_{q}}{H_{p}}(dy+\mathbf{A})^{2}
-\frac{\Delta_{\theta}}{H_{q}}(dt+\mathbf{B})^{2}+H_{p}\left(\frac{dr^{2}}{\Delta}+d\theta^{2}
+\frac{\Delta}{\Delta_{\theta}}\sin^{2}{\theta}d\phi^{2}\right)\,,
\end{equation}
where
\begin{eqnarray}
H_{p} & = & r^{2}+\alpha^{2}\cos^{2}{\theta}+r(p-2m)+\frac{p}{p+q}\frac{(p-2m)(q-2m)}{2} 
\nonumber\\
& & +\frac{p}{2m(p+q)}\sqrt{(q^{2}-4m^{2})(p^{2}-4m^{2})}\;\alpha\cos{\theta}\,, 
\label{eq:H1}\\
H_{q} & = & r^{2}+\alpha^{2}\cos^{2}{\theta}+r(q-2m)+\frac{q}{p+q}\frac{(p-2m)(q-2m)}{2} 
\nonumber\\
& & -\frac{q}{2m(p+q)}\sqrt{(q^{2}-4m^{2})(p^{2}-4m^{2})}\;\alpha\cos{\theta}\,, 
\label{eq:H2}\\
\Delta_{\theta} & = & r^{2}+\alpha^{2}\cos^{2}{\theta}-2mr\,, \label{eq:H3}\\
\Delta & = & r^{2}+\alpha^{2}-2mr\,, \label{eq:Delta}\\
\mathbf{A} & = & -\left[2Q(r+\frac{p-2m}{2})
-\sqrt{\frac{q^{3}(p^{2}-4m^{2})}{4m^{2}(p+q)}}\;\alpha\cos{\theta}\right]H_{q}^{-1}dt-
\Bigg[2P(H_{q}+\alpha^{2}\sin^{2}{\theta})\cos{\theta}\Bigg. \nonumber\\
& & -\sqrt{\frac{p(q^{2}-4m^{2})}{4m^{2}(p+q)^{3}}}[(p+q)(pr-m(p-2m))
+q(p^{2}-4m^{2})]\alpha\sin^{2}{\theta}\Bigg.\Bigg]H_{q}^{-1}d\phi\,, \label{eq:A}\\
\mathbf{B} & = & \sqrt{pq}\frac{(pq+4m^{2})r-m(p-2m)(q-2m)}{2m(p+q)\Delta_{\theta}}
\;\alpha\sin^{2}{\theta}d\phi\,. \label{eq:B}
\end{eqnarray}
The (four-dimensional) physical parameters are
\begin{eqnarray}
2G_{4}M & = & \frac{p+q}{2}\,, \label{eq:M} \\
G_{4}J & = & \frac{\sqrt{pq}(pq+4m^{2})}{4(p+q)}\frac{\alpha}{m}\,, \label{eq:J} \\
Q^{2} & = & \frac{q(q^{2}-4m^{2})}{4(p+q)}\,, \label{eq:Q} \\
P^{2} & = & \frac{p(p^{2}-4m^{2})}{4(p+q)}\,. \label{eq:P}
\end{eqnarray}
Solutions with black hole horizons have $q\geq 2m$, $p\geq 2m$, $m\geq |\alpha|$.

{}For regularity, the coordinate $y$ must be periodically identified as 
\beq\label{yperiod}
y\sim y+ 2\pi R\,,\qquad R=\frac{4 P}{N_6}\,,
\eeq
for integer $N_6$. As usual, $\phi\sim \phi+2\pi$. From a
five-dimensional viewpoint the KK electric charge is momentum
along the $y$-direction and so is quantized as
\begin{equation}\label{eq:quant_Q}
Q=\frac{2G_4N_{0}}{R}\,
\end{equation}
for integer $N_0$. In the string theory embedding, $N_0$ and $N_6$
correspond to the numbers of D0 and D6 branes introduced in \reef{d0d6},
and $R=g$ in string units.

We take a limit where the magnetic charge $P$ grows to infinity while
$Q$, $J$ and the black hole size remain finite. This has the effect of
effectively decompactifying the fifth direction. To perform this, we send
$p\to \infty$, and also send $r,m,\alpha,q\to 0$, and $y\to\infty$ in
such a way that $p\,r$, $p\,m$, $p\,\alpha$, $p\,q$, $y/p$, remain finite.
It is convenient to introduce new finite parameters $\mu$, $a$, $b$, and
finite radial and angular coordinates, $\rho$ and $\psi$, through
\begin{eqnarray}
p\,q & = & \frac{\mu}{4}\,, \label{eq:q_4} \\
p\,\alpha & = & \frac{1}{8}\left(\mu-(a+b)^{2}\right)^{1/2}(a-b)\,, \label{eq:alpha_4}\\
p\,m & = & \frac{1}{8}\left[\mu(\mu-(a+b)^{2})\right]^{1/2}\,, \label{eq:m_4}\\
p\,r & = & \frac{1}{4}\left[\rho^{2}-\frac{1}{2}
\left(\mu-a^{2}-b^{2}-\sqrt{\mu(\mu-(a+b)^{2})}\right)\right]\,, \label{eq:radial_4}
\end{eqnarray}
\beq\label{psiperiod}
\psi=\frac{y}{p}\,,\qquad \mathrm{with}\;\; \psi\sim\psi+\frac{4\pi}{N_6}\,.
\eeq
The angles $(\psi,\phi,\theta)$ are Euler angles
for (a topological) $S^3/\bbz{N_6}$. It is also convenient to use
\begin{equation}\label{eq:Euler}
\tilde{\psi}=\frac{\psi+\phi}{2}\,,\qquad  \tilde{\phi}=\frac{\psi-\phi}{2}\,,\qquad 
\tilde{\theta}=\frac{\theta}{2}\,,
\end{equation}
with
\beq
(\tilde{\psi},\tilde{\phi})\sim
\left(\tilde{\psi}+\frac{2\pi}{N_6},\tilde{\phi}+\frac{2\pi}{N_6}\right)
\sim (\tilde{\psi},\tilde{\phi}+2\pi)\,.
\eeq

After lengthy algebra, in the limit $p\to \infty$ the metric \reef{eq:general_KK} becomes
\begin{eqnarray}
ds^{2} & = & -dt^{2}+\frac{\mu}{\Sigma}\left(dt-a\sin^{2}{\tilde{\theta}}d\tilde{\psi}-
b\cos^{2}{\tilde{\theta}}d\tilde{\phi}\right)^{2}+\Sigma\left(\frac{d\rho^{2}}{\tilde{\Delta}}
+d\tilde{\theta}^{2}\right) 
\nonumber\\
& & +(\rho^{2}+a^{2})\sin^{2}{\tilde{\theta}}d\tilde{\psi}^{2}+
(\rho^{2}+b^{2})\cos^{2}{\tilde{\theta}}d\tilde{\phi}^{2}\,, \label{eq:general_MP}
\end{eqnarray}
with
\begin{eqnarray}
\Sigma & = & \rho^{2}+a^{2}\cos^{2}{\tilde{\theta}}+b^{2}\sin^{2}{\tilde{\theta}}\,, 
\label{eq:Sigma}\\
\tilde{\Delta} & = & \frac{(\rho^{2}+a^{2})(\rho^{2}+b^{2})-\mu \rho^{2}}{\rho^{2}}\,. 
\label{eq:Delta_MP}
\end{eqnarray}
This is the general five-dimensional MP black hole, with independent
rotation parameters $a$ and $b$ \cite{MP}.
When $N_6>1$ the orbifold identification \reef{psiperiod} implies that
the solution is not globally asymptotically flat, but instead the
spatial geometry asymptotes to $\bbr{4}/\bbz{N_6}$. In this case the MP
black hole sits at the tip of a conical space.

\subsection{Relations between physical parameters}

The 4D and
5D Newton constants are as usual related by
\beq 
G_{4}=\frac{G_{5}}{2\pi R}\,, \label{eq:R} 
\eeq 
with $R$ given in \reef{yperiod}.

The 4D mass, given by (\ref{eq:M}), is dominated in the limit
$p\to\infty$ by the magnetic KK monopole mass. This is identified as
\begin{equation}\label{eq:monopoleM}
 \mathcal{M}_{KK}=\frac{P}{2G_{4}}\,,
\end{equation}
and diverges as $p\to\infty$. The finite limiting difference between
the total 4D mass and the KK monopole mass corresponds exactly to the 5D mass,
\begin{equation}\label{5DM}
M-\mathcal{M}_{KK}\to \frac{3\pi}{8N_{6}G_{5}}\mu=M_{(5)}\,.
\end{equation}
According to this equation, we can regard the 5D mass as the excitation
energy above the KK monopole background. The $N_6$ in the denominator
comes from integration over the $\bbz{N_6}$-orbifolded $S^3$. 

Consider the following two Killing vectors of the KK black hole geometry,
\beq
\zeta_{(1),(2)}=2P\partial_y \pm \partial_\phi\,.
\eeq
Their associated conserved charges are\footnote{Our definitions of
$J_{1,2}$ differ from \cite{eh} by a factor of $N_6$.}
\beq\label{j12}
J_{1,2}=\frac{PQ}{G_4}\pm J=\frac{N_0N_6}{2}\pm J\,,
\eeq
which are independent of $R$ and therefore remain invariant as
$p\to\infty$. In this limit
\beq
\zeta_{(1)}\to \partial_{\tilde{\psi}}\,,\qquad \zeta_{(2)}\to
\partial_{\tilde{\phi}}\,,
\eeq
so $J_1$ and $J_2$ become the angular momenta of the MP black hole in
the directions
$\tilde{\psi}$ and $\tilde{\phi}$,
\beq
J_1\to\frac{\pi\mu a}{4G_5 N_6}\,,\qquad J_2\to\frac{\pi\mu b}{4G_5 N_6}\,.
\eeq
So, from \reef{j12}, the electric charge, which
is the component of the rotation aligned with the KK direction $y$,
corresponds to the 5D self-dual angular momentum $\cal J$
\begin{equation}\label{eq:N0}
N_{0}=\frac{J_{1}+J_{2}}{N_{6}}=\frac{\cal J}{N_{6}}\,,
\end{equation}
and the 4D angular momentum,
\begin{equation}\label{eq:anti_J}
J=\frac{J_{1}-J_{2}}{2}=\frac{\bar{\cal J}}{2}\,,
\end{equation}
is the 5D anti-self-dual angular momentum $\bar{\cal J}$. This is a $U(1)$ charge in the
$SU(2)\subset SO(4)/\bbz{N_6}$ that remains unbroken by the
compactification. It is the component of the rotation of the MP black
hole that lies away from the compactification direction. 

It can also be checked, with some work, that the entropy measured from
the area in four dimensions agrees in the limit with the
entropy of the MP black hole,
\begin{eqnarray}
S & = &\frac{\mathcal{A}_\mathrm{KKbh}}{4G_4}=
\frac{\pi\sqrt{pq}}{G_{4}}\left[m+\frac{pq+4m^{2}}{2m(p+q)}\sqrt{m^{2}-
\alpha^{2}}\right] \nonumber\\
&\to&
\frac{2\pi^{2}}{4G_{5}N_{6}}\mu\rho_{+}
=\frac{\mathcal{A}_\mathrm{MPbh}}{4G_5}
\end{eqnarray}
(with $\rho_+$ the outer horizon radius).

\subsection{The two extremal limits}

There are two different extremal limits for KK black holes, which
correspond to two different classes of extremal limit for the MP black
holes:

\begin{itemize}

\item Slowly-rotating extremal KK black holes are the limit of
\reef{eq:general_KK} where $\alpha,m\to
0$ with finite $|\alpha|/m<1$. This implies $G_4|J|<|PQ|$. 

In the
decompactification limit to the MP black hole, this corresponds to
\beq
\mu=(a+b)^2\quad \mathrm{with}\;\; ab>0\,, 
\eeq
which is
\beq\label{slowm5}
M_{(5)}^3=\frac{27\pi}{32 G_5 N_6}{\cal J}^2\,,\quad
\mathrm{with}\;\;|{\cal J}|>|\bar {\cal J}|\,.
\eeq
\item Fastly-rotating extremal KK black holes have $|\alpha|=m>0$, so
$G_4|J|>|PQ|$. In the decompactification
limit this is 
\beq
\mu=(a-b)^2\quad \mathrm{with}\;\; ab<0\,, 
\eeq
\ie
\beq\label{fastm5}
M_{(5)}^3=\frac{27\pi}{32 G_5 N_6}\bar {\cal J}^2\,,\quad
\mathrm{with}\;\;|\bar {\cal J}|>|{\cal J}|\,.
\eeq

\end{itemize}
The mass bound \reef{mgbound} translates into a bound on $M_{(5)}$ in
terms of $\cal J$,
\beq
M_{(5)}^3=\frac{27\pi}{32 G_5 N_6}{\cal J}^2\,,
\eeq
which is obviously saturated at extremal slow-rotation, \reef{slowm5},
and never saturated at fast rotation, \reef{fastm5}. From a purely 5D
viewpoint the distinction is arbitrary. It is only when we put the
solution at a certain orientation within Taub-NUT that the symmetry
between $\cal J$ and $\bar {\cal J}$ is broken.

Finally, since
\beq
J_1 J_2 =\frac{(N_0 N_6)^2}{4}-J^2\,,
\eeq
we can write the entropy as
\beq
S=2\pi \sqrt{|J_1 J_2|}\,,
\eeq
independently of $R$. The two extremal cases above correspond to
$J_1J_2>0$ and $J_1J_2<0$, respectively.

\setcounter{equation}{0}

\section{Ergospheres and superradiance in extremal KK-Black Holes}
\label{app:ergo}

Let us first consider a general necessary condition
for superradiant scattering, which follows from the second law of black
hole thermodynamics. From the 4D point of view, we must have
\begin{equation}\label{eq:first_law}
T_{H}dS=dM-\Phi_{E}dQ-\Phi_{M}dP-\Omega_H dJ>0
\end{equation}
(the condition is still valid in the extremal limit where $T_H\to 0$).
We only consider processes where the topology of the 5D solution remains
fixed, so $dP=0$. 

Consider a scalar field $\Psi$ in the black hole background \reef{eq:general_KK},
satisfying
\begin{equation}\label{boxpsi}
\square_{(5)}\Psi=0
\end{equation}
with the form 
\beq
\Psi=\psi_{\omega k n}(r,\theta) e^{iky+in\phi-i\omega t}\,.
\eeq
Here $k$ is interpreted as KK electric charge and is quantized in units
of $1/R$. From a 4D viewpoint it also gives a rest mass, so if the charged wave
is to propagate to infinity it must satisfy $\omega>|k|$.

It can easily be shown that absorption of this field by the black hole results
in a change in black hole parameters such that
\begin{equation} 
\frac{\delta J}{\delta M}=\frac{n}{\omega}\,,\qquad
\frac{\delta Q}{\delta M}=2G_4\frac{k}{\omega} \,.
\end{equation}
Then \reef{eq:first_law} requires
\begin{equation}\label{eq:first_law2}
\delta M\left(1-\frac{n}{\omega}\Omega_{H}-2G_4\frac{k}{\omega}
\Phi_{E}\right)>0\,.
\end{equation}
Since we are considering a process of energy extraction, $\delta M<0$,
the only way for this to hold is that
\begin{equation}\label{eq:general_superradiant}
\omega<n\Omega_{H}+2G_4 k\,\Phi_{E}\,.
\end{equation}

\begin{itemize}

\item For the slowly-rotating extremal solution the
four dimensional horizon has $\Omega_H=0$ and hence what we have is a
charge ergosphere. We can only extract energy by discharging the
black hole. The electric potential for these black holes is
\beq
2G_4\Phi_{E}=\sqrt{\frac{p+q}{q}}>1
\eeq
so the charge-superradiance condition can be satisfied.

\item The fastly-rotating extremal solution
has both non-zero angular velocity and electric potential on the
horizon,
\beq\label{omegah}
\Omega_H=\frac{1}{\sqrt{pq}}\,,
\eeq
and
\begin{equation}\label{eq:electric_potential}
2G_4\Phi_{E}=\sqrt{\frac{q^2-4m^2}{q(p+q)}}<1\,.
\end{equation}
Rotational superradiance of
neutral ($k=0$) waves is obviously possible. Bearing
in mind that $\omega>k$ for a
wave to escape, then it is not possible to extract energy
by simply discharging the black hole ($k>0,\,n=0$), but it seems possible to do
so by simultaneous extraction of angular momentum and charge.

\end{itemize}

Note that \reef{eq:general_superradiant} is necessary, but not
sufficient, for superradiance to be possible. Next we perform a detailed
analysis of scalar wave propagation to find, in particular illustrative
cases, that superradiance indeed happens when this is satisfied.
We consider extremal black holes with non-zero magnetic
charge, but set to zero either $J$ or $Q$, to obtain simple examples of
slowly and fastly rotating black holes.

\subsection{$Q\neq 0$, $J=0$ extremal black hole}

This is the static (in 4D) limit of slowly rotating black holes, obtained taking
$\alpha=0$ and then $m\to 0$. The horizon is at $r=0$ and from a 5D
viewpoint it is moving
along $y$. Indeed, the horizon is generated by orbits of the
Killing vector
\begin{equation}\label{eq:under_killing}
\xi=\frac{\partial}{\partial t}+v_{H}\frac{\partial}{\partial y}
\end{equation}
where
\begin{equation}\label{eq:under_vh}
v_{H}=\sqrt{\frac{p+q}{q}}\,.
\end{equation}
So the horizon is rotating at velocity $v_{H}$ relative to asymptotic
static observers that follow orbits of $\partial_{t}$. The vector
$\partial_{t}$ becomes spacelike for
$r<r_{e}=\frac{1}{2}\left(q-p+\sqrt{q^{2}+p^{2}}\right)$, so there is an
ergosphere, which from the 4D viewpoint is a charge ergosphere. The
velocity $v_H$
is actually the same as the KK electric potential $2G_4\Phi_{E}$. The
fact that $v_H>1$ does not result in any causal
pathology.

We now analyze if there are massless scalar superradiant modes in this
background. The
equation \reef{boxpsi} is separable for the \emph{ansatz}
\begin{equation}\label{eq:ansatz}
\Psi=\frac{f(r)}{\chi(r)}\Theta(\theta)e^{in\phi+iky-i\omega t}
\end{equation}
where
\begin{equation}\label{eq:under_chi}
\chi(r)=\left[\left(2qr^{2}+2pr(q+r)+p²(q+2r)\right)
\left(2qr(q+r)+p(q^{2}+2qr+2r^{2})\right)\right]^{1/4}\,.
\end{equation}
For the angular part we get
\begin{equation}\label{eq:polar}
\frac{1}{\sin{\theta}}\frac{d}{d\theta}\left(\sin{\theta}\frac{d\Theta}
{d\theta}\right)+\left(\lambda_{lnk}+
\frac{1}{\sin^{2}{\theta}}\left(pk\sqrt{\frac{p}{p+q}}+n\cos{\theta}\right)^{2}\right)
\Theta=0
\end{equation}
where $\lambda_{lnk}$ is a separation constant. 

For the radial part we obtain
\begin{equation}\label{eq:radial}
\frac{d^{2}f}{dr_{*}^{2}}+V(r)f=0
\end{equation}
where we have defined the `tortoise' radial coordinate $r_*$ as
\begin{equation}\label{eq:under_tortuise}
\frac{dr_{*}}{dr}=\frac{1}{2r^{2}(p+q)}\left(p^{3}q^{3}+4p^{2}q^{2}(p+q)r
+6pq(p+q)^{2}r^{2}+4(p+q)^{3}r^{3}+4(p+q)^{2}r^{4}\right)^{1/2}
\end{equation}
and whose asymptotic behaviour is
\begin{equation}\label{eq:asymp_tortuise}
\left\{
\begin{array}{lcl}
r_{*}\sim r & \mathrm{for} & r\to\infty\\
r_{*}\sim -\frac{1}{r} & \mathrm{for} & r\to 0\,\,.
\end{array}
\right.
\end{equation}
For the analysis of superradiance, we follow the approach of \cite{staro}, which only
requires the asymptotic behavior of
(\ref{eq:radial}).
Near the horizon
\begin{equation}\label{eq:horizon_potential}
V(r)\simeq\omega_{H}^{2}+O(r)\qquad(r\to0)
\end{equation}
with
\begin{equation}\label{eq:omegaH}
\omega_{H}=\omega-v_{H}k\,.
\end{equation}
Near infinity
\begin{equation}\label{eq:infinity_potential}
V(r)\simeq\omega_{\infty}^{2}+O(1/r)\qquad(r\to\infty)\,,
\end{equation}
where $\omega_{\infty}^{2}=\omega^{2}-k^{2}$. 
Then
\begin{equation}\label{eq:asymptotic_solution}
f(r)\sim\left\{
\begin{array}{ll}
e^{- i \omega_{\infty} r_{*}}+R e^{ i \omega_{\infty} r_{*}}\,, & r\to\infty \\
T e^{- i \omega_{H}r_{*}}\,, & r\to0\,,
\end{array}
\right.
\end{equation}
is a wave of unitary amplitude travelling
from infinity and then splitting into a transmitted wave of amplitude
$T$ that goes into the horizon, and a reflected wave of amplitude $R$
which goes back to infinity. 
If (\ref{eq:asymptotic_solution}) corresponds to a solution
of (\ref{eq:radial}) so does its complex conjugate
\begin{equation}\label{eq:asymptotic_solution2}
f^{*}(r)\sim\left\{
\begin{array}{ll}
e^{ i \omega_{\infty} r_{*}}+R^{*} e^{- i \omega_{\infty} r_{*}}\,, & r\to\infty \\
T^{*} e^{ i \omega_{H}r_{*}}\,, & r\to0\,.
\end{array}
\right.
\end{equation}

These two solutions are linearly independent, and the theory of
ordinary differential equations tells us that their wronskian
$W=ff'^{*}-f^{*}f'$ must be independent of $r$. Near
infinity this is $W=2 i \omega_{\infty}(|R|^{2}-1)$, and near
the horizon  $W=-2 i \omega_{H}|T|^{2}$. Equating
these we get
\begin{equation}\label{eq:wronskian_constant}
|R|^{2}=1-\frac{\omega_{H}}{\omega_{\infty}}|T|^{2}\,.
\end{equation}
A wave travelling from and to infinity ($\omega>|k|$) will undergo
superradiant amplification ($|R|>1$) if $\omega_H<0$, \ie
\begin{equation}\label{eq:under_superradiant}
k<\omega<v_{H}k\,.
\end{equation}
This can be always fulfilled since $v_{H}>1$. Since these black holes
have $\Omega_H=0$, it reproduces correctly the condition
\reef{eq:general_superradiant}.

\subsection{$Q=0$, $J\neq 0$ extremal black hole}

This is a particular case of extremal fastly rotating black holes.
It is obtained taking $|\alpha|=m$ and $q=2m$. The latter sets $Q=0$.

Now the horizon is at $r=m$. The
ergosphere is a squashed sphere given by $r=m(1+\sin{\theta})$,
which touches the horizon at $\theta=0$. The Killing horizon generator is
\begin{equation}\label{eq:over_killing}
\xi=\frac{\partial}{\partial t}+\Omega_{H}\frac{\partial}{\partial\phi}
\end{equation}
where
\begin{equation}\label{eq:over_omegah}
\Omega_{H}=\frac{1}{\sqrt{2mp}}\,.
\end{equation}

The procedure to study superradiance is as in the previous section. For
simplicity we consider a field without any dependence on $y$, so in 4D terms
this is an electrically neutral ($k=0$) scalar field. Since the background is
also neutral, the only effect of having $k\neq 0$ would be to give a 4D
mass to the field.

The Klein-Gordon equation is separable for an \emph{ansatz} of the form 
\begin{equation}\label{eq:ansatz2}
\Psi=\frac{f(r)}{\chi(r)}\Theta(\theta)e^{in\phi-i\omega t}
\end{equation}
with
\begin{equation}\label{eq:over_chi}
\chi(r)=\left[\left(m^{2}+r^{2}\right)\left(m^{2}-2mr+r(p+r)\right)\right]^{1/4}\,.
\end{equation}
The angular equation is 
\begin{equation}\label{eq:polar2}
\frac{1}{\sin{\theta}}\frac{d}{d\theta}\left(\sin{\theta}
\frac{d\Theta}{d\theta}\right)+\left(\lambda_{ln}+m^{2}\omega^{2}\cos^{2}{\theta}-
\frac{n^{2}}{\sin^{2}{\theta}}\right)\Theta=0
\end{equation}
which is the equation for spheroidal harmonics. The radial equation takes
again the form (\ref{eq:radial}), but with a different $V(r)$
and a different `tortoise' coordinate, now defined as
\begin{equation}\label{eq:over_tortuise}
\frac{dr_{*}}{dr}=\frac{1}{(r-m)^{2}}\left[\left(m^{2}+r^{2}\right)
\left(m^{2}-2mr+r(p+r)\right)\right]^{1/2}
\end{equation}
and with the asymptotic behavior 
\begin{equation}\label{eq:asymp_tortuise2}
\left\{
\begin{array}{lcl}
r_{*}\sim r & \mathrm{for} & r\to\infty\\
r_{*}\sim -\frac{1}{r-m} & \mathrm{for} & r\to m\,.
\end{array}
\right.
\end{equation}
The potential goes as
\begin{equation}\label{eq:over_asymp_V}
V(r)\to\left\{
\begin{array}{lcl}
\omega^{2}& \mathrm{for} & r\to\infty \\
(\omega-\Omega_{H}n)^{2}& \mathrm{for} & r\to m\,.
\end{array}\right.
\end{equation}
In this case we have no potential barrier near infinity from
the KK massess. Arguing as before, we get superradiant modes for
\begin{equation}\label{eq:over_superradiant}
0<\omega<\Omega_{H}n\,.
\end{equation}

\subsection{$Q\neq0$, $J\neq 0$, fastly rotating extremal black hole}

The generalization from the previous section to fastly rotating black
holes with $Q\ne0$ ($\alpha=m$, $q>2m$) is straightforward but very
cumbersome, so we do not provide here the full calculation. The result
one obtains is the natural generalization of the electrically neutral
case: (\ref{eq:over_superradiant}) is still valid, now with the general
value \reef{omegah} for $\Omega_H$. Since we are considering neutral
fields, $k=0$, this is in perfect agreement with
\reef{eq:general_superradiant}.

\vfill


 \end{document}